\documentclass[a4paper]{jpconf}
\usepackage{graphicx}
\begin{document}
\title{Theoretical developments in supernova neutrino physics : mass corrections and pairing correlators}

\author{Cristina Volpe}

\address{Astro-Particule et Cosmologie (APC), CNRS UMR 7164, Universit\'e Denis Diderot,\\ 10, rue Alice Domon et L\'eonie Duquet, 75205 Paris Cedex 13, France}

\ead{volpe@apc.univ-paris7.fr}

\begin{abstract}
We highlight the progress in our understanding of how neutrinos change their flavor in astrophysical environments, in particular effects from the neutrino self-interaction.
We emphasize extended descriptions of neutrino propagation in massive stars that are beyond the current one based on the mean-field approximation. The extended equations include, in particular, corrections from (anti)neutrino-(anti)neutrino pairing  correlations and from the neutrino mass. We underline open issues and challenges.

\end{abstract}

\section{Introduction}
\subsection{Core-collapse supernova neutrinos and SN1987A}
Supernovae (SNe) type II and Ib/c are massive stars that undergo gravitational collapse of their cores. Type II exhibits H lines in their spectra, type Ib has He and Si lines. SNe Ic show none of these indicating that before collapse the star has lost both the H envelope and He shell.  
The fate of a massive star is mainly determined by the initial mass, composition and the history of its mass loss. The explosion produces either a neutron star or a black hole, directly or by fallback. It was early realized that a gravitational binding energy of the order of $E \approx G M_{NS}^2/R_{NS} > 10^{53}$ erg associated with the core collapse to a neutron star (NS) would be released as neutrino emission which would eventually play a role in the ejection of the stellar mantle \cite{Colgate:1966ax}. 
Initial stellar masses range from 8 to 300 $M_{sun}$, $M_{sun}$ being the Sun's mass. Stars with lower masses develop an O-Ne-Mg core, while those with masses larger than about 9-10 $M_{sun}$  reach an iron core. Collapsars collapse to a black hole with an accretion disk. Low energy neutrinos are also emitted from this disk  or in binary neutron star mergers.

On 23 February 1987 Sk -$69^{\circ} 202$ exploded producing SN1987A, the first naked-eye supernova since Kepler's one in 1604. It was located in the Large Magellanic Cloud, a satellite galaxy of the Milky Way. The determined distance is 50 kpc from the Earth based on the expanding photosphere method. From the observed light-curve and simulations it appears that the core mass of SN1987A progenitor was around 6 $M_{sun}$, the total mass $\approx$ 18 $M_{sun}$ and the progenitor radius  about $10^{12}$ cm.  SN1987A is unique : observed in all wavelengths from gamma rays to radio, and for the first time, in neutrinos from the collapse of the stellar core. These neutrinos were first discovered  by Kamiokande II  (16 events) \cite{Hirata:1987hu}, then by IMB  (8 events) \cite{Bionta:1987qt} and Baksan (5 events) \cite{Alekseev:1988gp}. Several hours before, 5 events (often discarded in the analysis of SN1987A data) were seen in LSD detector  \cite{Aglietta:1987it}. 
The earliest observations of optical brightening were recorded 3 hours  after neutrino's arrival. There is currently no sign of its remnant like a bright pulsar as the one  from the supernova in the Crab nebula in 1054.

The basic features of core-collapse supernova predictions are confirmed by SN1987A events both concerning the neutrino fluence (time-integrated flux) and on their spectra.  A comparative analysis of the events gives as a best fit point $E = 5 \times 10^{52}$ ergs for the total gravitational energy radiated in electron anti-neutrinos and $T = 4$ MeV for their temperature \cite{Vissani:2014doa}. According to expectations, 99 $\%$ of the  gravitational binding energy should be converted in $\nu_e, \nu_{\mu}, \nu_{\tau}$ neutrinos (and anti-neutrinos) in the several tens of MeV energy range. Assuming energy equipartition among the flavors  one gets about $3 \times 10^{53}$ ergs. On the other hand the average electron anti-neutrino energy is 12 MeV at the best fit point if the neutrino spectra are to a fairly good approximation thermal. The emission time found is $15$ s  \cite{Vissani:2014doa}. Importantly SN1987A events have favored the "delayed" over the "prompt explosion mechanism"  \cite{Loredo:2001rx}.

Various mechanisms for the SN blast are investigated, including a thermonuclear, a bounce-shock, a neutrino-heating,  a magnetohydrodynamic, an acoustic and a phase-transition mechanisms (see  e.g.\cite{Janka:2012wk}). Since the kinetic energy in SN events goes from $10^{50-51}$ ergs for SNe up to several $10^{52}$ ergs for hyper-novae, the explosion driving mechanism have to comply, among others, with providing such energies. The neutrino-heating mechanism with non-radial hydrodynamical instabilities (convective overturn with SASI) appear to be a good candidate to drive iron-core collapse supernova explosions; while the more energetic hypernovae events could be driven by the magnetohydrodynamical mechanism. Note that a new neutrino-hydrodynamical instability  termed LESA  (Lepton-number Emission Self-sustained Asymmetry) has been identified  \cite{Tamborra:2014aua}.
Successful explosions  for two-dimensional (2D) simulations with realistic neutrino transport have been obtained. Simulations based on 3D are giving first results (see e.g. \cite{Melson:2015tia}).

The SuperNova Early Warning System (SNEWS) and numerous other neutrino detectors around the world can serve as supernova neutrino observatories if a supernova blows up in the Milky Way, or outside our galaxy. In particular, neutrinos would help locating the supernova, even if not visible, thanks to the directionality of neutrino-electron scattering, or using inverse-beta decay in running and future scintillator detectors \cite{Fischer:2015oma}.  
Large scale detectors based on different technologies including liquid argon, water Cherenkov and scintillator are on the way, in particular JUNO  \cite{An:2015jdp} and hopefully Hyper-K. These and the EGADS project (Super-Kamiokande with addition of Gadolinium) have the potential to detect neutrinos from an (extra)galactic  explosion and observe the diffuse supernova neutrino background from supernova explosions up to cosmological redshift of about 2.

 \section{Self-interaction effects in massive stars} 
 The intrinsic many-body nature of the neutrino evolution in presence of neutrino-neutrino interactions was already emphasised in \cite{Pantaleone:1992eq} where it was first pointed out that such interactions introduce a non-linear refractive index. 
In the last decade the study of neutrino self-interaction effects with mean-field equations has revealed a rich phenomenology of neutrino flavour conversion phenomena in dense media. 
Most studies are realised in a core-collapse supernova "set-up", while "collapsars"  or  neutron star-neutron star mergers also show interesting features at variance with the supernova case \cite{Malkus:2012ts}. In the "bulb model" flavour evolution occurs because of the bipolar instability, due to the divergence of the matter phase \cite{Galais:2011jh}. 
Flavour conversion goes beyond the Mikheev-Smirnov-Wolfenstein effect \cite{Wolfenstein:1977ue,Mikheev:1986gs}, encompassing e.g. magnetic resonance phenomena such as the spectral split \cite{Galais:2011gh}.   
For supernovae several schematic models have nowadays been investigated (sometimes using a linearisation of the equations of motion). Different kind of instabilities are found
in a range of parameters' values  (see e.g. \cite{Chakraborty:2015tfa}). In particular, a recent calculation for an inhomogeneous medium and including explicitly
time presents flavour changes in a region of neutrino and matter densities similar to those behind shocks in a supernova \cite{Dasgupta:2015iia}. 
Moreover  studies show that neutrino flavour conversion impacts nucleosynthetic abundances in supernovae, collapsars and neutron star mergers. 
Further work is needed to definitely assess the role of flavour conversion in these contexts and for future observations.
The effects of increasing dimensionality in the problem is an important open issue that is being investigated.
A key question is also the potential role of corrections beyond the mean-field since
current investigations are based on the mean-field approximation. 
In order to address this issue extended descriptions have been obtained.

 \section{Extended mean-field description of neutrino propagation : pairing correlators and mass contributions} 
In the past two decades different theoretical approaches have been employed to derive neutrino equations of motion in astrophysical environments. They can be classified as : {\it i)}  mean-field and extended mean-field equations ; {\it ii)} Boltzmann equations (see \cite{Volpe:2015rla} for a review). Note, that the Boltzmann equations used for the neutrino transport in core-collapse supernova simulations do not include the mixings and mean-field terms important for flavour evolution.  
In the most general mean-field equations, two-point correlation functions naturally arise that are associated with: {\it i)} non-zero neutrino masses \cite{Vlasenko:2013fja,Serreau:2014cfa} or neutrino magnetic moments \cite{deGouvea:2012hg,Studenikin:2004bu,Dvornikov:2011dv}; {\it ii)} neutrino-antineutrino pairing correlations \cite{Volpe:2013uxl,Vaananen:2013qja,Serreau:2014cfa}. 

A compact and simple quantum field theory derivation is given in Ref.\cite{Serreau:2014cfa} including both corrections from the mass and from pairing correlators.
This work  furnishes the most general mean-field equations for a massive Dirac or Majorana neutrino propagating in an inhomogeneous medium. One starts with all possible two-point correlators\footnote{The creation and annihilation operators $a, a^{\dagger}$ for neutrino and $b, b^{\dagger}$ for antineutrinos satisfy the canonical anti commutation rules.} : normal densities $ \rho_{\vec{p}',h',\vec{p},h}= \langle a^{\dagger}_{\vec{p},h} a_{\vec{p}',h'} \rangle$ for $\nu$, and $\bar{\rho}_{\vec{p}',h',\vec{p},h}  = \langle b^{\dagger}_{\vec{p}',h'} b_{\vec{p},h} \rangle $ for $\bar{\nu}$;  the abnormal densities $\kappa_{\vec{p}',h',\vec{p},h} =\langle b^{\dagger}_{\vec{p},h} a_{\vec{p}',h'} \rangle$ and $\kappa^{\dagger}$ ($\vec{p},h$ are momentum and helicity). Then the Ehrenfest theorem is applied with the most general mean-field Hamiltonian. In the Majorana case lepton-number violating $\kappa$ correlators are considered. 
The structure of the equations is given without need to specify the kernel that depends on the particles composing the medium and their interactions with the propagating neutrino. 
The results can be cast in matrix form: 
\begin{equation}\label{e:matrixform}
i\, \dot{\!{\cal R}} (t)= \left[ {\cal H}(t),{\cal R}(t)\right],
\end{equation} 
where the generalised Hamiltonian is 
\begin{equation}\label{e:genH}
{\cal H} (t)= \left(
\begin{array}{cc}   
\Gamma^{\nu\nu}(t) & \Gamma^{\nu\bar\nu}(t) \\
\Gamma^{\bar\nu\nu}(t)  & \Gamma^{\bar\nu\bar\nu}(t)  \end{array}
\right),
{\cal R}(t) =\left(
\begin{array}{cc}   
 \rho(t) &  \kappa(t) \\
\kappa^{\dagger}(t) &  1 - \bar{\rho}(t) \end{array}
\right),
\end{equation}
with $\Gamma^{\nu\nu}$ and $\Gamma^{\bar\nu\bar\nu}$ being the mean-field for neutrinos and antineutrinos respectively. The off-diagonal mean-field $\Gamma^{\nu\bar\nu}$ introduces a coupling  between neutrino and anti-neutrino sectors. Obviously,
in absence of mass contributions and pairing correlators Eq.(\ref{e:matrixform}) reduces to the Liouville von-Neumann equation, i.e. $i \dot{\rho}(t)  =  [h,\rho] $ (and similarly for anti-neutrinos)\footnote{The mean-field $h$ usually receives contributions from the mixings, the matter and the neutrino self-interaction (details can be found e.g. in \cite{Serreau:2014cfa,Volpe:2013uxl}).}, for the one-body density matrix $\rho$, in agreement with previous results (see e.g. Ref.\cite{Sigl:1992fn}).

In presence of mass contributions only ($\kappa, \kappa^{\dagger} = 0$), ${\cal H}(t)$ has a helicity structure with $ 2 \cal{N} \times$  2$\cal{N}$ elements, $\cal{N}$ being the number of neutrino families. For the Dirac case this gives for the neutrino sector
\begin{equation}
\label{eq:Hache}
 \Gamma^{\nu\nu}(t,\vec q\,)\!\to\!
   \left(\!\begin{tabular}{cc}
 $\Gamma^{\nu\nu}_{--}(t,\vec q\,)$&$\Gamma^{\nu\nu}_{-+}(t,\vec q\,)$\\
 $\Gamma^{\nu\nu}_{+-}(t,\vec q\,)$&$\Gamma^{\nu\nu}_{++}(t,\vec q\,)$
\end{tabular}\!  \right)\!\equiv\!
   \left(\!\!\begin{tabular}{cc}
 $H(t,\vec q\,)$&$\Phi(t,\vec q\,)$\\
 $\Phi^\dagger(t,\vec q\,)$&$\tilde H(t,\vec q\,)$
\end{tabular}\! \! \right)
\end{equation}
and a similar structure for the antineutrino mean-field $ \Gamma^{\bar\nu\bar\nu}$. The neutrino density matrix is
\begin{equation}
\label{eq:helicity1}
\rho(t,\vec q\,) \rightarrow
 \left(\begin{tabular}{cc}
 $\rho_{--}(t,\vec q\,)$&$\rho_{-+}(t,\vec q\,)$\\
 $\rho_{+-}(t,\vec q\,)$&$\rho_{++}(t,\vec q\,)$
 \end{tabular}
 \right) ,
\end{equation}
and similarly for $\bar{\rho}$.
The mass corrections contribute with, in particular, diagonal mean-field terms and the off-diagonal one  $ \Phi(t,\vec q\,)=e^{i\phi_q}\hat\epsilon_q^*\cdot\vec V(t){m\over 2q}, $ that are suppressed as expected by $m/2q$ with $m$ being the neutrino mass, $\hat\epsilon_q^*$ is a unit vector pointing to a direction perpendicular to the neutrino motion. $\vec V(t)$ is the vector component of the mean-field that depends on neutrino and matter anisotropic contributions. 
To be non-zero $\Phi(t,\vec q\,)$, termed  {\it spin} or {\it helicity coherence}, requires anisotropy of the medium and couples the neutrino and anti-neutrino sectors in the Majorana case, as pointed out in \cite{Vlasenko:2013fja}, or active with sterile components in the Dirac case \cite{Serreau:2014cfa}.
For the neutrino mass corrections, the study of a one-flavour model finds the conditions for a MSW-like resonance  and a non-linear feedback 
(under appropriate parameter choices) producing significant neutrino-antineutrino conversion \cite{Vlasenko:2014bva}. 

Possible contributions from $\kappa, \kappa^{\dagger}$ have been discarded so far on the basis that they oscillate very fast around zero (see e.g.\cite{Sigl:1992fn}). 
In presence of pairing correlators the equations of motions are given by Eqs.(\ref{e:matrixform}-\ref{e:genH}) where mean-field $\Gamma^{\nu\nu}, \Gamma^{\nu\bar\nu}, \Gamma^{\bar\nu\bar\nu}(t) $ terms are derived in Ref.\cite{Serreau:2014cfa}. Using the Born-Bogoliubov-Green-Kirkwood-Yvon hierarchy, Ref.\cite{Volpe:2013uxl} has provided for the first time extended mean-field evolution equations including contributions from neutrino-antineutrino pairing correlations. Such correlations are two-body corrections to the commonly used one-body density matrix description.  For homogeneous media, the off-diagonal term $\Gamma^{\nu\bar\nu}$ Eq.(\ref{e:genH}) requires medium anisotropy for the pairing correlators to contribute. Furthermore  the homogeneity condition for the background implies\footnote{As known, in the mean-field approximation for homogeneous environments the momentum of the propagating neutrinos is not changed.}, in particular, that $\kappa$ and $\kappa^{\dagger}$ involve  opposite momenta $(\vec{p}, \vec{p}')=(\vec{p}, - \vec{p})$. This condition is relaxed in inhomogenous media and momenta $\vec{p} + \vec{p}' \neq 0$ are correlated. 

Equations derived in Refs.\cite{Serreau:2014cfa,Volpe:2013uxl} show the presence of a source term implying that the usual argument to discard contributions from pairing correlators might not necessarily hold. In particular, such a term is sourced by the neutrino densities $\rho$ and $\bar{\rho}$ \cite{Serreau:2014cfa}.\footnote{The source term creates particle-antiparticle pairs. The effects of the created pairs do not introduce divergences because of the natural cutoff furnished by the scale of anisotropies and inhomogeneities in the system.} Therefore a non-zero $\kappa$ can develop in time even if it is zero initially. An {\it ansatz} on the size of the pairing correlations consists in assuming that at initial time the system is found in an eigenstate of the extend Hamiltonian with pairing correlations. A linearised analysis\footnote{Note that the linearised analysis corresponds to a small amplitude approximation and does not catch instabilities that arise in the large amplitude motion.} can be performed then around such an initial condition. This analysis can be done by introducing a generalised Bogoliubov-Valatin transformation and by going to a quasi-particle picture as shown in \cite{Vaananen:2013qja}, and exploited in \cite{Kartavtsev:2015eva}. 
However one should keep in mind that neutrino-neutrino, antineutrino-antineutrino or neutrino-antineutrino pairing correlations could be build up dynamically through, e.g. the collision term, as found in the context of leptogenesis and baryogenesis scenarios \cite{Fidler:2011yq}.

Contributions beyond the mean-field can play a role in the transition region between the high matter density opaque to neutrinos and the dilute regime where mean-field applies. 
This transition region is close to the neutrino-sphere, behind  the shock and potentially important for the explosion dynamics. 
Obviously contributions both from pairing correlators and mass contributions are expected to be small. However an amplification of their effects can occur due to the non-linearity of the evolution equations.  As for the role of collisions, Ref.\cite{Cherry:2012zw} has shown that a few collisions outside the neutrino-sphere can significantly influence flavour evolution. Realistic calculations of the Boltzmann equation for particles with mixings, provided e.g. \cite{Sigl:1992fn,Vlasenko:2013fja} are required to assess the competition among collisions, flavour and the macroscopic evolution of the exploding star. Such calculations represent one of the main challenges for the future.

The impact of pairing correlators on the neutrino evolution has not been determined yet.  In Ref.\cite{Kartavtsev:2015eva}\footnote{Note that in the Majorana case, a supplementary scalar term should also be added to the equations provided in Ref.\cite{Serreau:2014cfa}, as pointed out in Ref.\cite{Kartavtsev:2015eva}.} a simple estimate for an homogeneous background is provided to determine
if  a MSW-like resonance condition is met. This requires the off-diagonal term in the neutrino Hamiltonian to be of the same size as the difference of the diagonal terms which are of the order of twice the neutrino kinetic term.\footnote{The neutrino and anti-neutrino kinetic terms on the diagonal appear with opposite signs; while usually they are taken out of the diagonal because they are proportional to the identity matrix.} However, one needs to ascertain if non-linear effects can take place that are of a different nature than the MSW effect. In inhomogeneous backgrounds the situation might be different since pairing correlators do not involve particles with back-to-back momenta  \cite{Volpe:2013uxl,Serreau:2014cfa}. 
Finally the results of Ref.\cite{Dasgupta:2015iia} show that large scale instabilities can be triggered by very small fluctuations. These fluctuations could be provided by pairing correlations. Maybe the numerical findings in Ref.\cite{Dasgupta:2015iia}  represent a first clue of the potential role of pairing correlations introduced in \cite{Volpe:2013uxl}. Clearly, we still need to address intriguing and challenging questions  to achieve a solid and in-depth understanding of neutrino flavour evolution and its implication in astrophysical environments.


\section*{References}

\begin{thebibliography}{9}

\bibitem{Colgate:1966ax} 
Colgate S A and White R H,
1966 {\it Astrophys.\ J.}  {\bf 143} 626.

\bibitem{Hirata:1987hu} 
Hirata K {\it et al.},
1987  {\it Phys.\ Rev.\ Lett.}  {\bf 58} 1490.

\bibitem{Bionta:1987qt} 
 Bionta R M, {\it et al.}, 
1987 {\it Phys.\ Rev.\ Lett.}  {\bf 58} 1494.


\bibitem{Alekseev:1988gp} 
  E.~N.~Alekseev, L.~N.~Alekseeva, I.~V.~Krivosheina and V.~I.~Volchenko,
1988  {\it Phys.\ Lett.} B {\bf 205} 209.


\bibitem{Aglietta:1987it} 
Aglietta M {\it et al.},
1987 {\it  Europhys.\ Lett.}  {\bf 3} 1315.



\bibitem{Vissani:2014doa} 
Vissani F,
2015  {\it J.\ Phys.\ G} {\bf 42} 013001.

\bibitem{Loredo:2001rx} 
Loredo T J and Lamb D Q,
2002 {\it   Phys.\ Rev.} D {\bf 65}, 063002.


\bibitem{Janka:2012wk} 
  Janka H T,
2012  {\it  Ann.\ Rev.\ Nucl.\ Part.\ Sci.}  {\bf 62} 407.
  
\bibitem{Tamborra:2014aua} 
 Tamborra I {\it et al.},
2014  {\it  Astrophys.\ J.}  {\bf 792} 96.


\bibitem{Melson:2015tia} 
Melson T, Janka H T and Marek A,
  arXiv:1501.01961 [astro-ph.SR].  




\bibitem{Fischer:2015oma} 
Fischer V {\it et al.},
 2015   {\it JCAP} 08 032..
 
\bibitem{An:2015jdp} 
  F.~An {\it et al.},
  arXiv:1507.05613 [physics.ins-det].
  
    
\bibitem{Pantaleone:1992eq} 
Pantaleone J T,
1992   {\it Phys.\ Lett.\ B} {\bf 287} 128.
 
 
\bibitem{Malkus:2012ts} 
Malkus A, Kneller J P, McLaughlin G C and Surman R,
2012  {\it  Phys.\ Rev.\ D} {\bf 86} 085015.
  
    
\bibitem{Galais:2011jh} 
  Galais S, Kneller J and Volpe C,
2012  {\it J.\ Phys.} G {\bf 39} 035201.
  
\bibitem{Wolfenstein:1977ue} 
Wolfenstein L.,
 1978  {\it Phys.\ Rev.\ D} {\bf 17} 2369

\bibitem{Mikheev:1986gs} 
Mikheev S P and Smirnov A Y,
1985  {\it  Sov.\ J.\ Nucl.\ Phys.} {\bf 42} 913
   [1985 {\it  Yad.\ Fiz.}  {\bf 42} 1441].

  
\bibitem{Galais:2011gh} 
  Galais S and Volpe C,
2011 {\it  Phys.\ Rev.} D {\bf 84} 085005.





\bibitem{Chakraborty:2015tfa} 
  Chakraborty S, Hansen R H, Izaguirre I and Raffelt G,
  arXiv:1507.07569 [hep-ph].

\bibitem{Dasgupta:2015iia} 
 Dasgupta B and Mirizzi A,
 2015  {\it   Phys.\ Rev.} D {\bf 92} 125030.
  
   
\bibitem{Volpe:2015rla} 
Volpe C,
 2015  {\it  Int.\ J.\ Mod.\ Phys.} E {\bf 24} no. 09 1541009.

\bibitem{Vlasenko:2013fja} 
Vlasenko A, Fuller G M  and Cirigliano V,
2014  {\it  Phys.\ Rev.\ D} {\bf 89}  105004.

\bibitem{Serreau:2014cfa} 
Serreau J and Volpe C, 2014,
{\it  Phys.\ Rev.\ D} {\bf 90} 125040.
   
\bibitem{deGouvea:2012hg} 
  De Gouvea A and Shalgar S, 
2012  {\it  JCAP }{\bf 1210} 027.
    
\bibitem{Studenikin:2004bu} 
  Studenikin A I,
 2004  {\it Phys.\ Atom.\ Nucl.}  {\bf 67} 993  [ 2004 {\it Yad.\ Fiz.} {\bf 67} 1014].

  
\bibitem{Dvornikov:2011dv} 
  Dvornikov M,
 2012 {\it Nucl.\ Phys.} B {\bf 855} 760.

  
\bibitem{Volpe:2013uxl} 
  Volpe C, Vaananen D and Espinoza C,
2013 {\it  Phys.\ Rev.\ D} {\bf 87}  113010.
 
\bibitem{Fidler:2011yq} 
Fidler, Herranen M, Kainulainen K and Rahkila P M,
2012  JHEP {\bf 1202} 065.

\bibitem{Sigl:1992fn} 
  Sigl G and Raffelt G,
1993  {\it     Nucl.\ Phys.\ B} {\bf 406} (1993).


\bibitem{Vlasenko:2014bva} 
Vlasenko A, Fuller G M and Cirigliano V,
  arXiv:1406.6724.

\bibitem{Cherry:2012zw} 
Cherry J F, Carlson J, Friedland A, Fuller G M and Vlasenko A,
2012 {\it  Phys.\ Rev.\ Lett.} {\bf 108} 261104.




  
  
\bibitem{Kartavtsev:2015eva} 
  Kartavtsev A, Raffelt G and Vogel H,
2015  {\it  Phys.\ Rev.} D {\bf 91} 125020.
  


\bibitem{Vaananen:2013qja} 
  V\"{a}\"{a}n\"{a}nen D and Volpe C,
2013  {\it   Phys.\ Rev.} D {\bf 88} 065003.

\end{thebibliography}
\end{document}